%% file: 0-main.tex
\documentclass[sigconf,,]{acmart}

\renewcommand\footnotetextcopyrightpermission[1]{} 
\newcommand{\etal}{\textit{et al. }}

\acmConference[]{}{}{}

\def\BibTeX{{\rm B\kern-.05em{\sc i\kern-.025em b}\kern-.08emT\kern-.1667em\lower.7ex\hbox{E}\kern-.125emX}}
    
\usepackage{mathtools}
\usepackage{amssymb,amsmath}
\usepackage{multirow}
\usepackage{tabularx}
\usepackage{graphicx}
\usepackage{subfig}
\usepackage{tcolorbox}
\usepackage{physics}
\usepackage{listings}
\usepackage{enumitem,kantlipsum}
\usepackage{balance}

\newcolumntype{N}{>{\raggedright\arraybackslash}X}
\newcolumntype{L}{>{\hsize=0.6\hsize\raggedright\arraybackslash}X}
\newcolumntype{R}{>{\hsize=0.6\hsize\raggedleft\arraybackslash}X}

\author{Maxime Cordy}
\affiliation{
  \institution{{  University of Luxembourg, Luxembourg}}
  }
\email{maxime.cordy@uni.lu}

\author{Renaud Rwemalika}
\affiliation{%
  \institution{{  University of Luxembourg, Luxembourg}}
}
\email{renaud.rwemalika@uni.lu}

\author{Mike Papadakis}
\affiliation{%
  \institution{{  University of Luxembourg, Luxembourg}}
 }
\email{michail.papadakis@uni.lu}

\author{Mark Harman}
\affiliation{%
  \institution{{ University College London, UK}}
  }
\email{mark.harman@ucl.ac.uk}

\begin{document}

\title[FlakiMe: Laboratory-Controlled Test Flakiness Impact Assessment. A Case Study on Mutation Testing and Program Repair]{FlakiMe: Laboratory-Controlled Test Flakiness \\ Impact Assessment. A Case Study on Mutation Testing and Automated Program Repair }


\begin{abstract}
Much research on software testing makes an implicit assumption that test failures are deterministic such that they always witness the presence of the same defects. However, this assumption is not always true because some test failures are due to so-called flaky tests, i.e., tests with non-deterministic outcomes. Unfortunately, flaky tests have major implications for testing and test-dependent activities such as mutation testing and automated program repair. To deal with this issue, we  introduce a test flakiness assessment and experimentation platform, called FlakiMe, that supports the seeding of a (controllable) degree of flakiness into the behaviour of a given test suite. Thereby, FlakiMe equips researchers with ways to investigate the impact of test flakiness on their techniques under laboratory-controlled conditions. 
We use FlakiME to report results and insights from case studies that assesses the impact of flakiness on mutation testing and program repair. These results indicate that a 5\% of flakiness failures is enough to affect the mutation score, but the effect size is modest (2\% - 4\% ), while it completely annihilates the ability of program repair to patch 50\% of the subject programs. We also observe that flakiness has case-specific effects, which mainly disrupts the repair of bugs that are covered by many tests. Moreover, we find that a minimal amount of user feedback is sufficient for alleviating the effects of flakiness. 
\end{abstract}

%


\maketitle 

\input{1-introduction.tex}
\input{2-relatedword.tex}
\input{3-rqs.tex}
\input{4-framework.tex}
\input{5-setup.tex}
\input{6-results.tex}
\input{7-conclusion.tex}

\balance
\bibliographystyle{ACM-Reference-Format}
\bibliography{bibliography,slice}


\end{document}

%% file: 1-introduction.tex
\section{Introduction}
\label{sec:introduction}

Test flakiness is the property of a test case and system under test, that the test can pass on one occasion, yet fail on another, without the tester changing anything other than the fact that the test is executed on two different occasions. This behaviour has a number of causes, such as nondeterminism in the system under test, instability in the infrastructure that provides the test environment, and variability in the results produced by services and components upon which the system under test depends.

Flakiness has a profound impact on all applications of software testing, because it increases test signal uncertainty; the tester can never be sure that a failure is genuine and this may waste effort (investigating false positives) or lose important signals (from switching off flaky tests). Companies such as Google and Facebook have highlighted the problem of test flakiness \cite{LeongSPTM19, memon:taming,mhpoh:scam18-keynote, Lam2019}, indicating that it is one of their primary concerns for software testing. Some companies have also launched specific challenges to the research community to tackle this problem \cite{facebook:TAVRFP2019}.

Flakiness impacts each form of testing in different ways. For example in mutation testing, the mutation score will vary, dependent on flakiness, confounding this variability with the influence of the quality of the test that the score seeks to assess. In program repair, the certainty we have that a repair is correct will be affected by flakiness, as will be the ability of the repair technique to localise the point at which to attempt a patch. Indeed, it has been argued that {\em all} forms of testing need to be reformulated to take account of flakiness to find techniques that can cope well in the presence of unavoidable flakiness \cite{mhpoh:scam18-keynote, AlshahwanCH0MMM19}. This means that testing techniques need to be re-investigated under flakiness conditions to assess their robustness on varying degrees of flakiness. 

In order to address the problems posed by flakiness, researchers need ways to investigate the impact of flakiness. Clearly work should be done on real world systems in the field to explore this impact \cite{LeongSPTM19, memon:automated}. However, researchers also need the ability to experiment with flakiness in laboratory controlled conditions. Such laboratory control would allow researchers to report results on the impact of {\em varying degrees} of flakiness on the test techniques they propose and introduce.

To address this need, we introduce a test flakiness tool, FlakiMe, that allows researchers to deliberately seed a (controllable) degree of flakiness into the behaviour of a given test suite and a given system under test. FlakiMe equips researchers with a laboratory controllable environment in which to experiment. This paper introduces the FlakiMe platform and illustrates its application to the assessment of flakiness impact on the problems of mutation testing and automated program repair.

Previous research on flakiness has helped to identify the main causes of flakiness \cite{Palomba2017a,memon:automated,luo:flaky}, and has introduced techniques to either reduce or ameliorated its effects. However, hitherto, no systematic way of evaluating the effect of flakiness on arbitrary software testing problems has been introduced. We fill this gap and report results on the use of our FlakiMe platform to yield insight on two software testing problems. 
Specifically, we perform a case study showing results for two cases of software testing:

\begin{enumerate}[leftmargin=0.4cm]
\item {\bf Mutation testing}: flakiness impacts the mutation score and we show how to investigate the {\em size} of this effect. 
In particular, using FlakiMe we can reveal that: 
(a) A small amount of flakiness affects mutation score (5\% of flakiness yields 2\% to 4\% mutation score variations); 
(b) As the degree of flakiness increases, a saturation point is quickly reached (approximately between $5\%$ and $10\%$ flakiness) after which further increases in the degree of flakiness have rapidly diminishing effects. 
(c) These first two findings, taken together, yield a take-home message for mutation testing researchers: flakiness is a potential problem for mutation testing but its effects are not that big. 
 Although researchers should always take into account the flakiness effects on the mutation scores they report, the results suggest that it is not sufficient to ``poison the well''.  

\item {\bf Automated program repair}:
For the application of repair, we found that the impact of flakiness is more profound than it is on mutation testing. Specifically we show that the same degree of flakiness has profoundly different effects on different systems. This indicates that research on automated repair needs to analyze how sensitive to flakiness their test suites and subjects are. Our results reveal that FlakiMe can be used to pre-select suitable subjects and to validate the key decisions made by the studied techniques. More precisely, we show that: 
    (a) Deterministic repair is increasingly affected by the number of tests covering the produced patches. Fortunately, we found that ``genuine'' patches are on average $12\%$ more likely to remain unaffected than non-genuine ones. 
    (b) Non-deterministic repair experiences a drop in the number of patches produced by $7\%$ to $100\%$, with the worst case (total failure; 100\% drop) occurring for 50\% of the programs studied. Exploiting knowledge about the non-flaky failures reduces this effect, yielding up to 4 times more patches and allowing successful repair on cases that flakiness previously impeded it.
\end{enumerate}


%% file: 2-relatedword.tex
\section{Related Work}
\label{sec:related}

Previous work on test flakiness \cite{Palomba2017a, Palomba2019, Lam2019, Presler-Marshall2019, Ahmad2019} has primarily focused on identifying its causes. Yet very few studies \cite{LeongSPTM19, Shi2019} analyze the effects of flakiness on software testing and test dependent techniques. As such, the primary goal of almost all previous work is to define the problem and understand the root causes of the non-determinism of the test signals. 

Luo \etal \cite{luo:flaky} proposed a formal classification of the root causes of test flakiness. The follow-up work aimed at the automated identification of flaky test \cite{Gao2016c} and the development of tools that remove flaky tests, such as iFixFlakies \cite{Shi2019}, iDFlakies \cite{Lam2019a}, RootFinder \cite{Lam2019}, DeFlaker \cite{Bell2018}. 

To the best of our knowledge, Shi \etal\cite{Shi2019} and Leong \etal\cite{LeongSPTM19} are the only ones investigating the effects of test flakiness on software testing techniques. The former study investigated the impact on mutation testing, while the later one investigated the impact on regression test selection. 

Shi \etal evaluated the effects of flakiness when computing mutation score and proposed a way to reduce this problem. The study shows that the mutation score can vary up to 5\% when ignoring the non-determinism of tests. These findings are based on \textit{in vivo} test flakiness not \textit{in vitro} laboratory controlled flakiness as we introduce here. Their findings are consistent with ours, suggesting that the laboratory control is well calibrated with in the wild real world flakiness.  

Leong \etal\cite{LeongSPTM19} investigated, at the Google CI environment, various test selection algorithms and report that flaky tests tend to significantly mislead their actual performance. 

Finally, in their keynote, Harman and O'Hearn \cite{mhpoh:scam18-keynote} highlighted the importance of adequately handling flaky tests and suggested that future research should ``Assume all Tests Are Flaky''. Therefore, their key suggestion is to develop `flakiness-robust' solutions capable of maximising the testing value in the presence of inherent or unavoidable flakiness. FlakiMe supports experiments with the above view by introducing controlled flaky behaviour on test suites.  



%% file: 3-rqs.tex
\section{Motivation and Objectives}
\label{sec:rqs}

To assess FlakiMe we investigate the impact of test flakiness on mutation testing and automated program repair. 

Mutation testing \cite{MutationSurvey} measures the strengths of test suites by determining the proportion of mutants (artificially injected defects) causing tests to transition from passing to failing.  
Hence, failed executions of flaky tests can artificially inflate the metric (mutation score) causing significant overestimation of the fault revealing potential 
of test suites. Thus, in our first question we examine (quantitatively) the extent to which mutation score can be inflated by flakiness:

\begin{description}
    \item[RQ1] \emph{To what extent does flakiness artificially inflate the mutation score of given test suites?}
\end{description}

To answer this question we check the effect of flakiness on the mutation scores of randomly chosen test suites (sampled from the projects' test suites). Our interest is on the divergence of those scores under different degrees of flakiness. 

Automated program repair aims at generating patches (modifications of the software code that fix bugs) for programs with bugs witnessed by failing test cases. In this line of work, effectiveness is measured by the number of valid patches (i.e., patches making all tests pass), generated within a given time limit. 
In some cases, researchers check these (valid) patches and label them as \emph{genuine} when they are considered as semantically equivalent (after manual inspection) to the real-world patch created by a developer (developer patches form the ground truth). Since the validity of patches is determined by the test results, it is interesting to see the extent to which flakiness can impact their selection. In other words, we would like to check the sensitivity of repair methods on flakiness. Hence, we ask:

\begin{description}
    \item[RQ2] \emph{To what extent does flakiness hinder the effectiveness of program repair at generating valid patches?}
\end{description}


To answer this question we select two recent repair methods, PRActical Program Repair (PRAPR)~\cite{Ghanbari2019} and Automated Repair for Java Programs (ARJA) \cite{Yuan2018}, that exhibit fundamental differences in the way they are working (PRAPR uses mutation testing, while ARJA uses genetic programming).  

PRAPR 
applies Fault Localization (FL) \cite{WongGLAW16} to associate an estimated degree of suspiciousness to the statements covered by the failing tests. 
It then ranks the statements according to their suspiciousness with the intention of increasing the likelihood of finding a good fixing point early.

To repair the programs, PRAPR applies a predefined set of mutations on the prioritized statements (according to the suspiciousness rank established by the FL). This process results in a set of patches, which are executed with all tests that cover the mutated statement (including the initially-failing tests). The patches that pass all tests constitute the resulting set of valid patches.

ARJA is a GenProg-like \cite{GouesNFW12} tool. It generates a population of patches that evolve over a predefined number of generations. ARJA first runs the whole test suite and applies FL on the failing tests. 
Then, it discards the statements with suspiciousness values below a predefined threshold and collects the statements that (i) are covered by at least one test covering the suspicious statements (ii) have some dependency with the suspicious statements. The collected statements form the set of \emph{ingredients}. A patch is formed by randomly altering the ingredients. ARJA uses the NSGA-II genetic algorithm to make the patches evolve over the generations. It works with a fixed-size population and keeps producing patches until a fixed number of patches have been evaluated. To evaluate a patch, ARJA runs the initially-failing tests and all other (passing) tests that cover suspicious statements (not discarded during filtering). 

A common characteristic of PRAPR and ARJA is that the total number of attempted patches remains constant over different runs. However, PRAPR does it in a deterministic way (for the same buggy program and failing tests, it generates systematically the same patches). Therefore, flakiness may cause valid patches to be categorised as invalid, thereby reducing the number of valid (and genuine) patches produced by the tool. On the contrary, ARJA produces patches randomly, guided by genetic programming. The number of generated valid patches is therefore also arbitrary and flakiness lowers this number. Based on these observations we divide RQ2 into the following subquestions:

    \emph{To what extent does flakiness decrease the number of valid/genuine patches produced by deterministic mutations?}\\

    \emph{To what extent does flakiness decrease the 
    number of valid patches produced by genetic programming?}\\

Going a step further we also investigate the way PRAPR and ARJA use FL. 
PRAPR applies FL only on the failing test cases while, ARJA runs all tests cases before applying FL (does not assume that the failing test cases are known). This implies that flakiness may increase the set of tests targeted by FL and, in turn, augment the set of ingredients (statements to mutate) with statements covered by the failing flaky tests. This has two consequences: (a) the search space encompasses more candidate patches that do not fix the bug (since they target wrong statements), reducing de facto the effectiveness of ARJA; (b) the number of tests -- both failing and passing -- executed to evaluate candidate patches is increased with tests covering the new ingredients. Hence, flakiness not only lowers the probability of ARJA to generate valid patches but it also assigns them a non-zero probability to be invalid (because of flaky test failures), thereby reducing the success rate of the tool. In view of this, we also investigate a slightly different scenario where the user specifies one (or some) failing test(s). In this case, the partial knowledge of some `real' failing tests could help alleviating the effects of flakiness. Therefore we ask:

\begin{description}
    \item[RQ3] \emph{Does making fault localization target real failing tests improve the robustness of program repair against flakiness?}
\end{description}

Finally, we also consider the way the two methods exploit the results of FL. PRAPR prioritizes the statements to mutate by ordering them according to suspiciousness but ultimately considers all statements, while ARJA discards statements whose degree of suspiciousness is below a predefined threshold. This difference can lead to a significant increase in the suspicious statements, leading again to an increased search space in the presence of flakiness. Additionally, this increase can reduce the suspiciousness of some statements, putting ARJA in a situation where real buggy statements are ignored. Thus, our last question concerns the sensitivity of FL on flakiness with respect to the suspicious statement selection:

\begin{description}
    \item[RQ4] \emph{How does flakiness affect the threshold-based suspicious statement selection by program repair techniques?}
\end{description}

Overall, our study aims at demonstrating that FlakiMe leads to interesting insights on the techniques' behaviour, when put under flakiness conditions. Our goal is to show that some decisions and methods' characteristics, which deserve attention, can be easily noticed through the lens of FlakiMe. We demonstrate how FlakiMe offers such opportunities.

%% file: 4-framework.tex
\section{FlakiMe}

FlakiMe 
injects flakiness on the results of test suites. As such it suggests that flakiness should be a parameter put under experimental control. FlakiMe implementation allows flaking JUnit tests seamlessly, with minimal changes to the test code and without modifying the program source code. It consists of dedicated JUnit runners extending the default runner, which works on a method-by-method (i.e. test by test) basis. 

\subsection{FlakiMe Test Runners}

The example code in Listing~\ref{lst:flakime} illustrates an implemented method of FlakiMe where each test method (test case) receives a 0.05 probability to fail. 
When failing due to flakiness it triggers an unchecked exception named \texttt{FlakiException}. Additional information is also recorded, in dedicated global variables \texttt{nbTests}, \texttt{nbPassed} and \texttt{nbFlaked} (respectively), i.e., the total number of test executed, the number of those that reach the end of their execution, and the number of those that flaked.

\begin{lstlisting}[float,language=java,caption={FlakiMe example implementation. Each passing test receives a 0.05 probability to flake. The number of tests that run, pass and flake are recorded in global variables.},label={lst:flakime}]
public class FlakiMe extends BlockJUnit4ClassRunner {
...
    protected final void runAtomic(Statement statement, Description description,
            RunNotifier notifier) {
        EachTestNotifier eachNotifier = new EachTestNotifier(notifier, description);
        eachNotifier.fireTestStarted();
        nbTests++;
        try {
            statement.evaluate();
            if (Math.Random() < 0.05) {
        	   eachNotifier.addFailure(new FlakiException());
        	   nbFlaked++;
            } else {
                nbPassed++;
            }
        } catch (AssumptionViolatedException e) {
            eachNotifier.addFailedAssumption(e);
        } catch (Throwable e) {
            eachNotifier.addFailure(e);
        } finally {
            eachNotifier.fireTestFinished();
        }
    }
}
\end{lstlisting}

FlakiMe can be tailored to different laboratory conditions controlling the occurrence of flakiness. For instance, one can change the effect of flakiness (making test transition from pass to fail, from fail to pass, or both) and its probability of occurrence (independent and uniformly for each test, dependent on the previous number of flaked test, dependent on whether tests execute similar or specific parts of code that flaked earlier or not, etc.).


To run tests with a FlakiMe runner, instead of JUnit's default runner, one should only add FlakiMe as a dependency on the program (e.g., using Maven) and use the \texttt{@RunWith} annotation on top of the test classes containing tests to flake. Additionally, by relying on runners one can easily apply different customized scenarios (implemented through different runners), reflecting different occurring conditions of flakiness, on different test classes. Even so, the fine-grained functioning of FlakiMe allows customizing these conditions for each test case. Runners are singletons -- they are instantiated only once for each run of the test suite -- which allows keeping a global view on all tests executed through each runner. Finally, FlakiMe can evolve independently of JUnit (as opposed to introducing flaking capabilities in a forked version of JUnit).

\subsection{FlakiMe: Mutation Testing}

Mutation testing generate mutants $M_1, \dots, M_k$ by altering the syntax of the original programs. It evaluates test suites strengths by running the tests with the mutants. The test suite $t_1, \dots, t_n$ kills a mutant if the execution of a test on this mutant fails (assuming that the execution passes on the original program). The mutation score (number of mutants killed divided by the number of mutants) is a frequently used metric for measuring test thoroughness \cite{MutationSurvey}. One can see this as an $n \times k$ matrix, where each cell is related to a test $t_i$ and mutant $M_j$ pair, and denotes whether $t_i$ killed $M_j$ or not. 

In the absence of flakiness such a matrix is determined by the tests and the mutants. However, in the presence of flakiness things change arbitrarily; a flaky test that passes on the original program can fail on a mutant leading to a kill (instead of mutant survival). Thus, running the test suite with FlakiMe results in different matrices, where the status of mutants for some test cases is swapped from survived to killed. The probability of swapping in this case is equivalent to the probability of the concerned tests to flake. 


\subsection{FlakiMe: Program Repair}

In program repair 
a valid patch is defined as one that compiles and passes all tests, including the initially failing tests (tests witnessing the bug). FlakiMe impacts this validity check by making a test fail arbitrarily. Based on the formed techniques, such a patch could be discarded although it should not. 

\subsubsection{Deterministic mutation-based repair (PRAPR)} Assume a buggy program with a non-flaky test suite including the failing tests. We denote by $P$ the set of patches generated by PRAPR on a given buggy program, by $V \subseteq P$ the set of valid patches, and by $G \subseteq V$ the set of genuine patches. The use of FlakiMe can introduce flaky test failures, which does not change $P$ but decreases $V$ and, thus, $G$. 
Accordingly, the probability $\overline{p_v}$ for a (initially) valid patch $v \in V$ to be labelled as invalid due to flakiness is the probability that any test $t_v$ covering $v$ flakes and fails. That is, $\overline{p_v} = P(\cup_{t_v \in T_v} fail_{t_v)})$ where $T_v$ is the set of tests covering $v$ and $fail_{t_v}$ denotes the event where test case $t_v$ fails because of flakiness. In the case where the occurrence of these events are independent and identically distributed with a failure probability $p$, running FlakiMe yields $\overline{p_v} = 1 - (1-p)^{T_v}$. In other words, valid patches have more risk to be wrongly labelled as invalid when the number of tests that cover them is higher. Accordingly, the expected number of valid (resp. genuine) patches in the presence of flakiness is $\mathbb{E}(|V_f|) = \sum_{v \in V} (1 - \overline{p_v})$ (resp. $\mathbb{E}(|G_f|) = \sum_{g \in G} (1 - \overline{p_g})$) and the probability to generate at least one valid (resp. genuine) is given by $P_v = 1 - (\prod_{v \in V} \overline{p_v})$ (resp. $P_g = 1 - (\prod_{g \in G} \overline{p_g})$). The assumptions behind the above analytic solutions make possible a laboratory-controlled test flakiness solution 
that can provide insights on the long-term behaviour of the test suite. 

\subsubsection{Genetic programming-based repair (ARJA)} ARJA generates the same number of candidate patches over different runs. However, the patches will differ due to the randomness in the evolution of the population. Hence, the number and content of valid patches varies from one run to another. Interestingly, even under some assumptions, analytic solutions aiming at computing this number are hard to set and not available. Therefore, the impact of FlakiMe can only be observed empirically.

Flaky tests may also impact the initial test suite run, impacting the fault localization estimates (suspiciousness scores). This can have a double effect; change the patch search space and alter the number of tests to be used for patch validity check. These effects can be observed on ARJA's report (number of detected failing tests and number of \emph{positive tests}, i.e., the passing tests that cover one or more suspicious statements). 
Variations in these numbers provide a coarse view on the extent to which the use of FlakiMe has reshaped the search space and affects the likelihood of finding a valid patch. 

\subsection{FlakiMe: Suspicious Statement Selection}

Suspicious statement selection in most repair techniques -- including PRAPR and ARJA -- is performed as follow. 
Given a set of statements $\{s_1, \dots, s_s\}$ and a test suite $\{t_1, \dots, t_n\}$, FL assigns a suspiciousness score to each statement based on the number of failing and passing tests covering them. It does this by building an $n \times s$ matrix where each cell records whether particular test covers a particular statement. Then, it runs all tests and keeps record of the tests that passed and failed. Based on this, it applies a similarity formula that assigns a suspiciousness score to each statement. For instance, Ochiai, the metric used by PRAPR and ARJA, assigns any statement $s$ to the score: $s_f/\sqrt{(s_f + n_f )\cdot (s_f + s_p)}$
where $s_f$ is the number of failing tests covering the statement $s$, $n_f$ is the number of failing tests and $s_p$ the number of passing tests covering $s$. 

Compared to a non-flaky test suite with clearly identified failing tests, FlakiMe affects the Ochiai score of all statements because tests sometimes fail instead of passing. This increases the values $n_f$ and $s_f$ (if the failing flaky tests cover $s$). In the end, flakiness can either increase or decrease the Ochiai score of the statements. When a surrounding repair method discards statements based on their suspiciousness score (as it is the case for ARJA), such differences can largely affect the search space and, thus, further reduce the effectiveness of the repair. 

%% file: 5-setup.tex
\section{Experimental Setup}
\label{sec:setup}


Mutation testing and program repair are typically evaluated on programs with passing tests. To observe how flakiness affects these techniques, we use FlakiMe to make tests that pass, fail in a non-deterministic way (injecting flakiness) at the end of their execution. The converse case (i.e. applying repair on failing tests that non-deterministically pass) would cause analogous situations and is therefore omitted from our experiments.

We also assume that the subject techniques have no prior knowledge on the causes of flakiness (e.g. which tests are flaky and what is the rate of occurrence of flakiness failures). Recent studies \cite{Labuschagne2017a,Bell2018,Ahmad2019,LeongSPTM19} highlight various degrees of flakiness in industrial code bases. We represent this degree as the ``flakiness failure rate'', which represents the average percentage of tests that are flaky and fail. This is a parameter that we control in our experiments. Thus, we assign, on every flaky test, the same probability of failure (i.e. flakiness failure rate).


\subsection{Tools}

We used the open-source tool PIT \cite{ColesLHPV16}, with its default operator set, to produce mutants and compute the mutation score. The mutation score $MS$ of test suite $T$ on a program $P$ can be expressed as: $MS(P,T)=\abs{K}/(\abs{M} - \abs{E})$
where $\abs{K}$ is the number of mutants killed, $\abs{M}$ is the total number of mutants and $\abs{E}$ is the number of equivalent mutants. We ignore equivalent mutants $E$ since they do not impact our analysis and thus we use a simplified mutation score measure: $\overline{MS}(P,T)=\abs{K}/\abs{M}$. Thus, as flakiness is introduced, only the number of killed mutants $\abs{K}$ influences the mutation score.

PRAPR is available as a Maven plugin and as a Docker image.\footnote{https://github.com/prapr/prapr} We used the Docker image to replicate the original experiments and retrieve the relevant measurements (e.g. number of tests that cover each mutant).

We also use the official implementation of ARJA retrieved from GitHub\footnote{\url{https://github.com/yyxhdy/arja}}, which we modified only to print additional statistics related to its execution. In our experiments, we run it on a MacBook Pro 2018 with macOS 10.14.5 and Java 1.7.0\_80-b15. To account for random variations in the patch generation process, we execute 10 runs of ARJA for each experiment. This repetition number was a compromise between statistical relevance and computation cost (one single run of ARJA on one of the subject programs can take more than 2 hours on our 2018 MacBook Pro 2.9 GHz Core-i7).

\subsection{Test Subjects}

Defects4J \cite{Just2014} is a set of real bugs harvested from Java projects. It is one of the most popular sets in evaluating program repair techniques, including PRAPR and ARJA. In our experiments we consider the bugs were the techniques succeeded. An important success metric here is the ability of the techniques to generate genuine patches (semantically-equivalent to the developers' patch). Thus, for PRAPR, we picked the 20 buggy projects for which PRAPR produced at least one genuine patch. We discarded the buggy programs for the Closure project because PRAPR requires more than 64GB of RAM to repair them \cite{Ghanbari2019}. For ARJA, we consider 8 of the buggy programs for which the tool generated at least one genuine patch (reported in ARJA's supplementary material\footnote{\url{https://github.com/yyxhdy/arja-supplemental/blob/master/arja-supplemental.pdf}}) and for which we could successfully generate valid patches (using the default settings of the tool). Unfortunately, we could not generate valid patches for some programs, probably due to differences in the tool configurations and/or infrastructures. Nevertheless, to increase diversity, we also considered 3 projects for which ARJA could generate valid (but not genuine) patches.

For mutation testing, we consider the latest releases (non-buggy) of the projects whereof we use buggy versions in the repair experiments. We choose these projects to maintain a certain consistency across our experiments. 

%% file: 6-results.tex
\section{Results}
\label{sec:results}

\subsection{RQ1: Mutation Testing}

\begin{figure*}
\vspace{-0.5em}
\centering
\subfloat[Mutation score]{\includegraphics[width=0.31\linewidth]{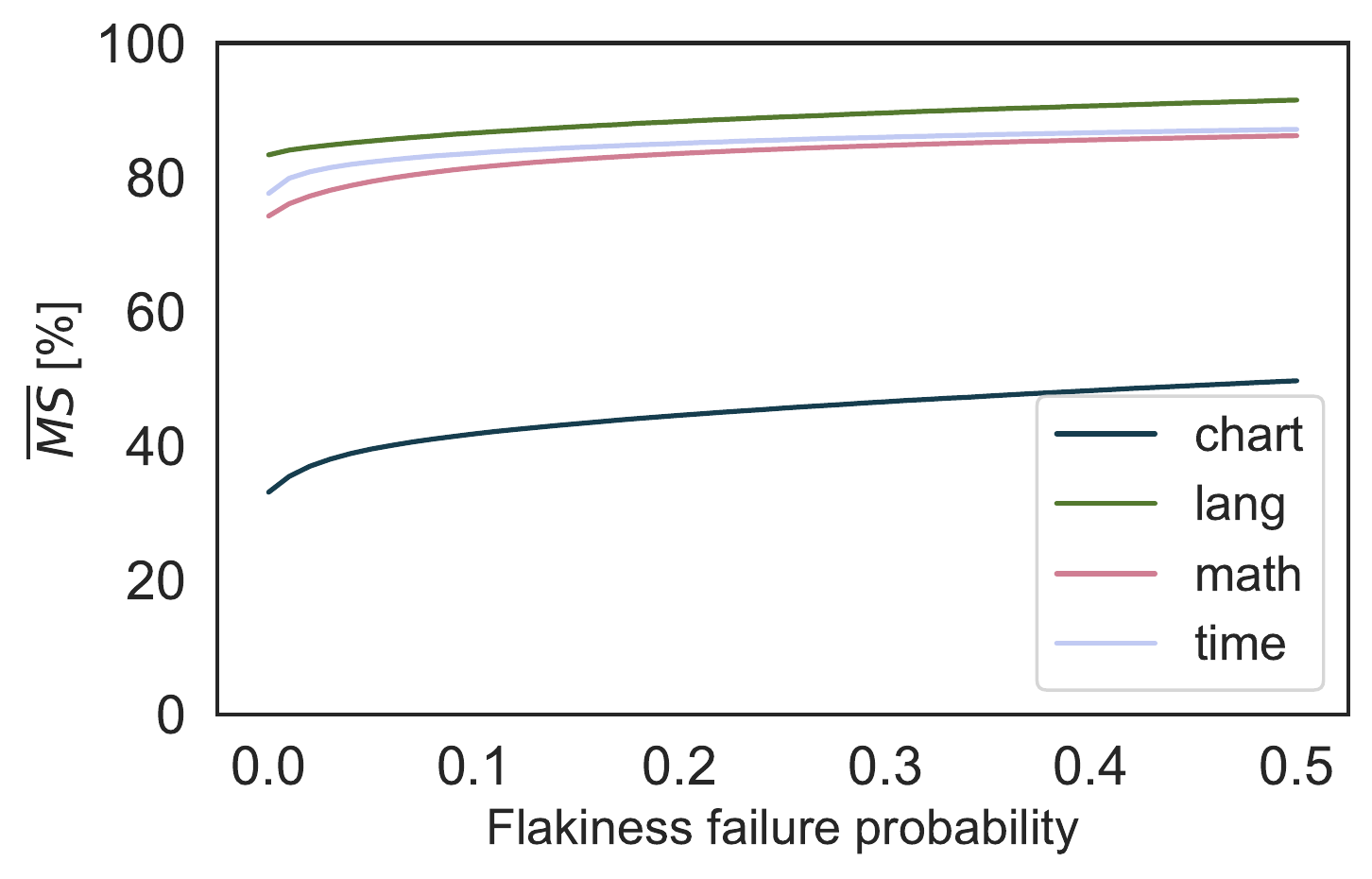} \label{fig:mutation_score}}
\subfloat[Standard deviation]{\includegraphics[width=0.31\linewidth]{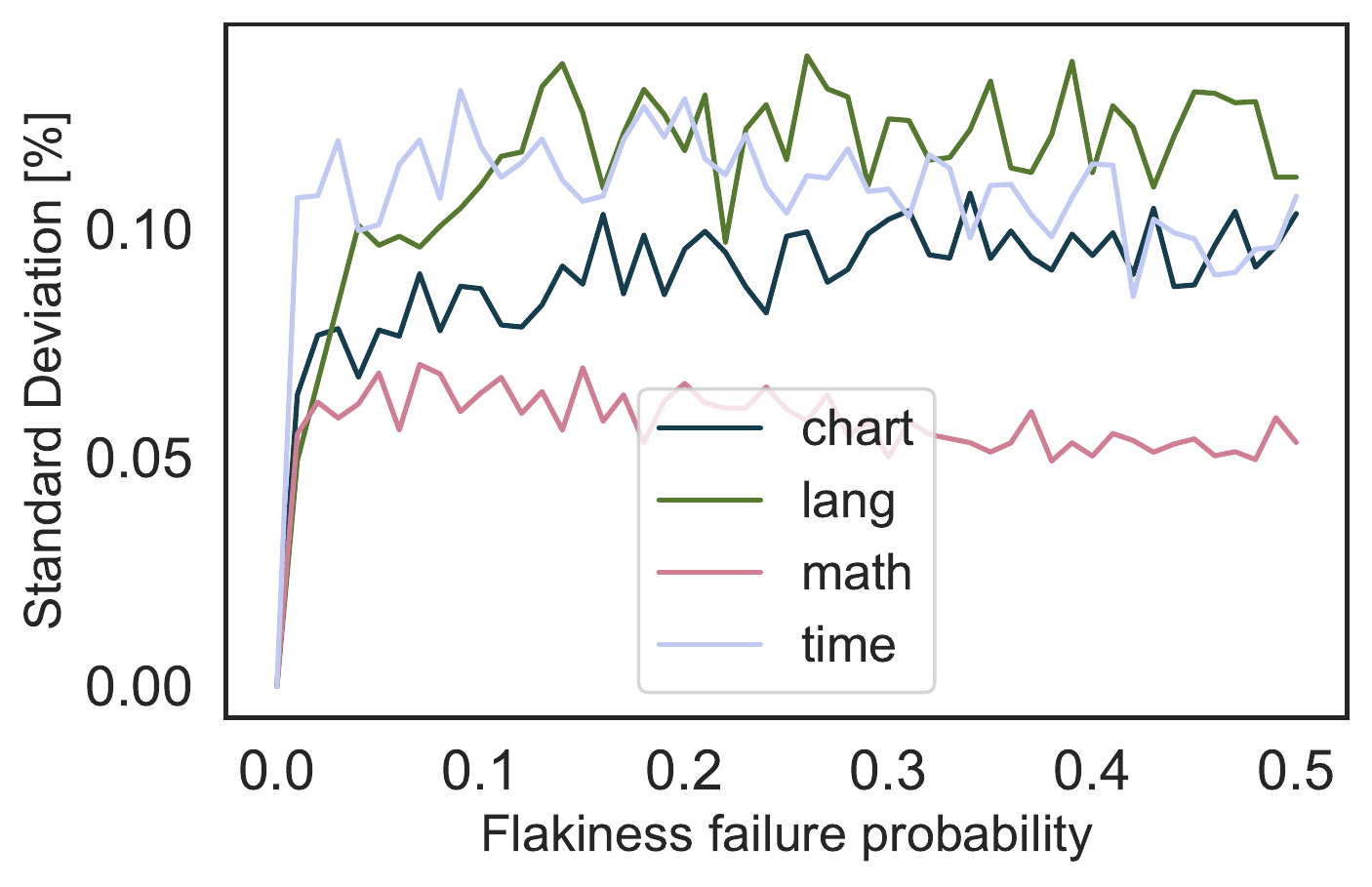} \label{fig:mutation_std}}
\subfloat[Difference]{\includegraphics[width=0.31\linewidth]{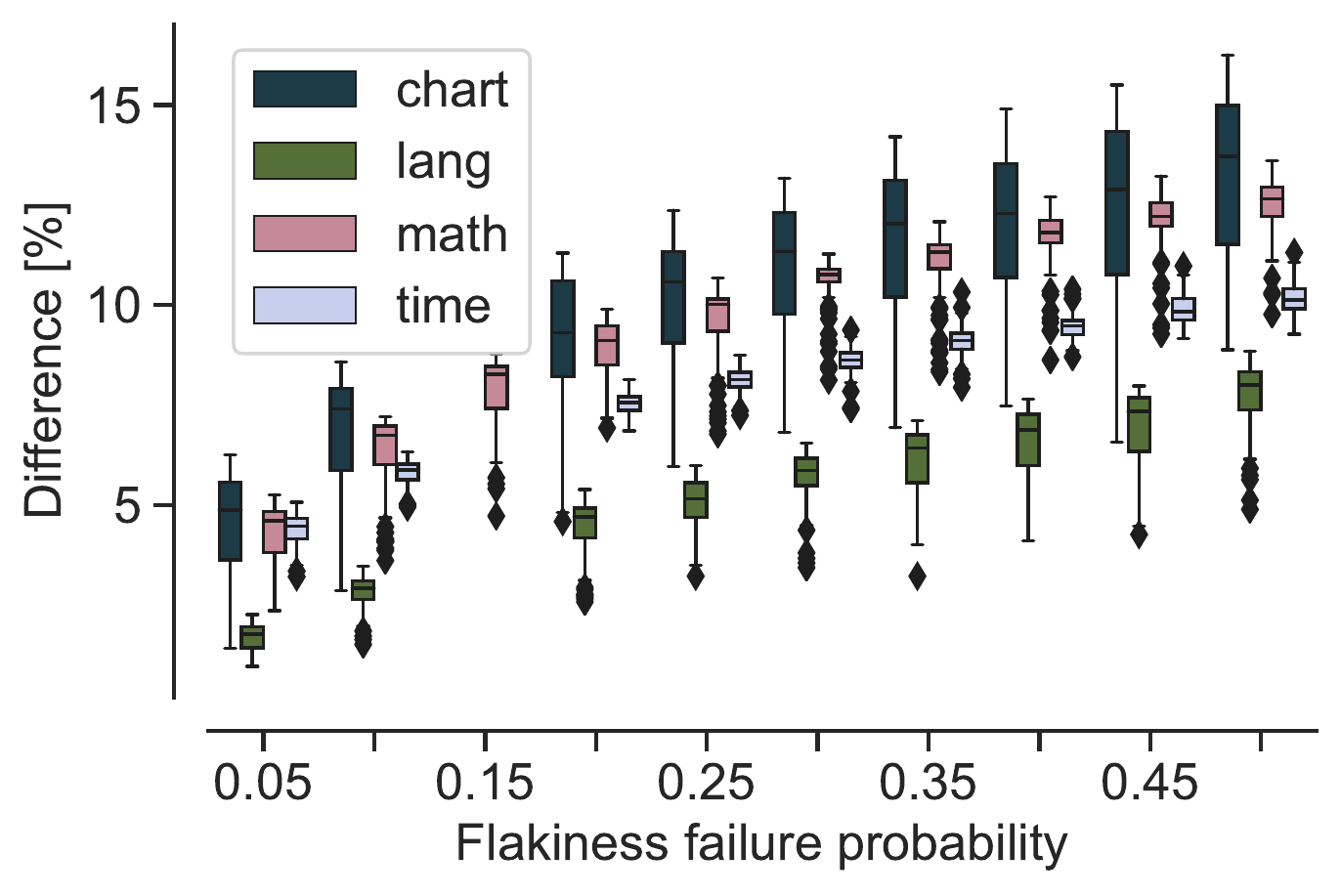} \label{fig:mutation_difference}}
\caption{Assessing the impact of the flakiness on the mutation score when flakiness failure level increases. Figure \ref{fig:mutation_score} shows the $\overline{MS}$, figure~\ref{fig:mutation_std} shows the standard deviations of $\overline{MS}$ and figure~\ref{fig:mutation_difference} shows the difference between the flaky $\overline{MS}$ and the non flaky one, when considering test suites of various sizes.\\ }
\vspace{-0.5em}
\label{fig:mutation}
\end{figure*}

We investigate the effect of flakiness (failure probabilities from 0 to 0.5, by step of 0.01) by running PIT 100 times for each probability. We analyze the variation of the mutation scores when the probability increases. Figure~\ref{fig:mutation_score} shows the results. We observe that mutation score increases more when the flakiness is at a low level. This indicates that the existence of flaky tests is enough to introduce noise even if flakiness is at a small level, but the effect is modest. This effect depends on the projects and is bounded by the number of survived mutants that are covered by the flaky tests (asymptotic behaviour). We also observe that projects with lower mutation score experience a greater score increase than those with higher scores. For instance, the mutation score of JFreeChart raises from 33.12\% to 49.71 while the score of Common Lang increases from 83.35\% to 91.51\%.

Figure~\ref{fig:mutation_std} shows the standard deviation of the mutation score when flakiness failure probability increases. When there is no flakiness, the outcome of the tests is deterministic and therefore, we observe a standard deviation of zero. When flakiness occurs, the standard deviation is low with average values ranging from 0.06\% (Commons Math) to 0.12\% (Commons Lang). 

To quantify the extent to which mutation score is inflated in a more general case, we randomly select test suites (by sampling from the original ones). The samples are of random size, ranging between 10\% and 90\% of the original test suite size. Doing so, allows observing the impact of flakiness on a more general and wide range of mutation score levels. Therefore, for every flakiness degree, we sampled 100 test suites (per program) and compute the mutation score differences between the non-flaky and flaky cases.

Figure~\ref{fig:mutation_difference} shows the flakiness effect on the randomly selected test suites. The boxes represent the differences in the mutation scores (flaky score minus non-flaky score) of 100 randomly selected test suites. From these results we can see that the median values for a flakiness probability of 0.05 are ranging between 1.76\% (Commons Lang) and 4.87\% (JFreeChart). Additionally, we see that even though the metric is disrupted, e.g., for JFreeChart, the difference on the scores ranges between 1.42\% and 6.25\%, the effect is moderate. Furthermore, we observe a small variation (higher 75\% quartiles - lower 25\% quartiles) between the test suites ranging from 2\% to 4\%. This indicates that the metric is relatively stable as there are only small relative differences between the test suites. \\

\begin{tcolorbox}[colback=white]
Flakiness inflates the mutation score but the effect is modest. 5\% of flakiness introduce approximately 2\%-4\% disruption on the metric. 
\end{tcolorbox}

\subsection{RQ2: Effectiveness of Program Repair}

\subsubsection{Deterministic technique.}

\begin{table}
\vspace{-1.5em}
    \caption{Impact of flakiness (with flakiness probability $0.05$) on the numbers of valid patches and genuine patches generated by PRAPR. $|P|$, $|V|$ and $|G|$ denote, respectively, the number of all patches, valid patches and genuine patches originally generated by PRAPR. \emph{CPP} (Covering Per Patch) is the average number of tests covering a valid patch. $\mathbb{E}(|V_f|)$ is the expected number of valid patches in the flaky case, whereas $P_v$ and $P_g$ are, respectively, the probability of generating at least one valid and at least one genuine patch. The capability of PRAPR to generate valid/genuine patches is reduced as more tests cover those patches. In some cases, flakiness annihilates any chance of generating valid patches.}
    \label{tab:results-prapr}
    
    \centering
    \begin{tabular}{|l|r|r|r|r|r|r|r|r|r|r|r|}
        \hline
         \textbf{Bug} & $|P|$ & $|V|$ & $|G|$ & CPP. & $\mathbb{E}(|V_f|)$ & $P_v$ & $P_g$ \\
         \hline 
            math-5 & 419 & 3 & 1 & 36.33 & 1.09 & .96 & .95\\
            math-34 & 258 & 1 & 1 & 2.00 & .90 & .90 & .90\\
            math-50 & 1138 & 40 & 1 & 3.10 & 34.48 & 1.00 & .70\\
            math-59 & 2417 & 1 & 1 & 1.00 & .95 & .95 & .95\\
            math-75 & 718 & 1 & 1 & 1.00 & .95 & .95 & .95\\
            math-82 & 2694 & 9 & 1 & 14.00 & 4.39 & 1.00 & .49\\
            math-85 & 1606 & 4 & 1 & 17.00 & 1.67 & .89 & .42\\
            
            time-11 & 3597 & 41 & 1 & 6.34 & 29.72 & 1.00 & .95\\
            time-19 & 4666 & 2 & 1 & 713.00 & .00 & .00 & .00\\
            
            lang-6 & 268 & 1 & 1 & 31.00 & .20 & .20 & .20\\
            lang-57 & 10 & 3 & 1 & 11.00 & 1.71 & .57 & .57\\
            lang-59 & 121 & 2 & 1 & 3.00 & 1.71 & .86 & .86\\
            
            mock.-29 & 3959 & 6 & 1 & 5.00 & 4.64 & .77 & .77\\
            mock.-38 & 510 & 3 & 1 & 77.67 & .43 & .14 & .21\\
            
            chart-1 & 3704 & 2 & 1 & 38.00 & .28 & .26 & .14\\
            chart-11 & 158 & 2 & 1 & 16.00 & .88 & .69 & .44\\
            chart-12 & 2245 & 2 & 1 & 3.50 & 1.67 & .97 & .81\\
            chart-20 & 240 & 1 & 1 & 95.00 & .01 & .01 & .01\\
            chart-24 & 133 & 2 & 1 & 1.00 & 1.90 & 1.00 & .95\\
            chart-26 & 12422 & 111 & 1 & 43.63 & 13.90 & 1.00 & .28\\
         \hline
    \end{tabular}
\vspace{-1.5em}
\end{table}

To evaluate the impact of flakiness on PRAPR, we first replicate its original experiments \cite{Ghanbari2019}. We retrieve the sets $P$, $V$ and $G$ (set of patches generated by PRAPR, set of valid patches, and set of genuine patches), as well as the sets $T_v$ (set of tests covering $v$) for each mutant $v \in V$. Then, assuming a flakiness failure rate of $5\%$, we set $p = 0.05$. We can then derive the expected number of valid patches ($\mathbb{E}(|V_f|)$) and genuine patches ($\mathbb{E}(|G_f|)$) in the presence of flakiness and compare with the original results. We also estimate the probability to generate at least one valid (resp. one genuine) patch.

Table \ref{tab:results-prapr} records, for every test subject, the impact of flakiness on the number of generated patches that are valid/genuine. Depending on how many tests cover the valid matches, the expected number of valid patches is reduced from $5\%$ to almost $100\%$, while the probability of generating at least one valid patch ranges from almost $0.00$ to almost $1.00$ (as apposed to the non-flaky case that always have a probability of $1.00$). In 14 cases out of 20, this probability is higher than 0.88, which shows that PRAPR often generates at least one valid patches covered by a small number of test cases. Math-5, for example, involves only 3 valid patches with a 36.33 average number of covering tests. However, $P_v$ remains as high as 0.96, which means that one of the patches is covered by very few tests. Math-50, Time-11 and Chart-26 are particularly comfortable: their larger number of valid patches leads to a high probability that at least one of them will make it through (no flaky test will flake). For some projects, however, flakiness has a disastrous impact: PRAPR generates only two and one valid patches for Time-19 and Chart-20, while their average number of covering tests is very high (713 and 95, respectively). 

In all cases, the use of PRAPR without any flakiness, generates only one genuine patch.\footnote{According to \cite{Ghanbari2019}, PRAPR often produces only one genuine patch, sometimes two.} Thus, in the flaky case we have $\mathbb{E}(|G_f|) = P_g$. PRAPR generates the unique genuine patch with a probability of almost $0.00$ to $0.95$. This solely depends on the tests covering the genuine patch and the probability of flakiness failures. The worst cases are, again, Time-19 and Chart-2 due to their high number of covering tests, while the best cases are those where the average number of covering tests is low. Interestingly, for some projects (e.g. Math-5, Time-11), PRAPR maintains a high probability (> 0.95) of generating the genuine patch even though the average number of covering tests is high. The genuine patch is actually covered by few tests (only one in the case of Math-5) compared to the other patches. More generally, $P_g$ is, on average, $12\%$ higher than the average probability for a valid patch to remain valid, which is given by $\mathbb{E}(|V_f|)/|V|$. Thus, when confronted with flakiness, PRAPR should first generate the valid patches covered by the smallest number of covering tests (thereby maintaining a low chance of failure due to flakiness) and check if one of them is genuine.

Based on these results we conclude that the ability of PRAPR to generate valid patches mainly depends on the scope of the bugs. This means that local bugs that are targeted by very few test cases (e.g. at the unit level) are more likely to be fixed correctly. On the contrary, bugs lying at the crossroad of the program's execution flows that are covered by many tests are, therefore, much harder to be successfully fixed by PRAPR. These results are aligned with previous studies showing that deterministic repair methods become less effective as the number of failing tests increases \cite{Kong2018}. This observation can also help researchers reduce the manual effort involved when verifying whether valid patches are genuine, by prioritizing towards patches covered by a smaller number of tests. \\

\begin{tcolorbox}[colback=white]
Flaky tests reduce the effectiveness of deterministic repair techniques by 5\% to 100\%. They decrease it more when a) fewer patches are generated and b) these patches are covered by more (potentially flaky) tests. Fortunately, the genuine patches we examined are covered by fewer tests than the non-genuine ones.
\end{tcolorbox}


\subsubsection{Non-deterministic technique.}

\begin{figure}
    \centering
    \includegraphics[width=0.75\linewidth]{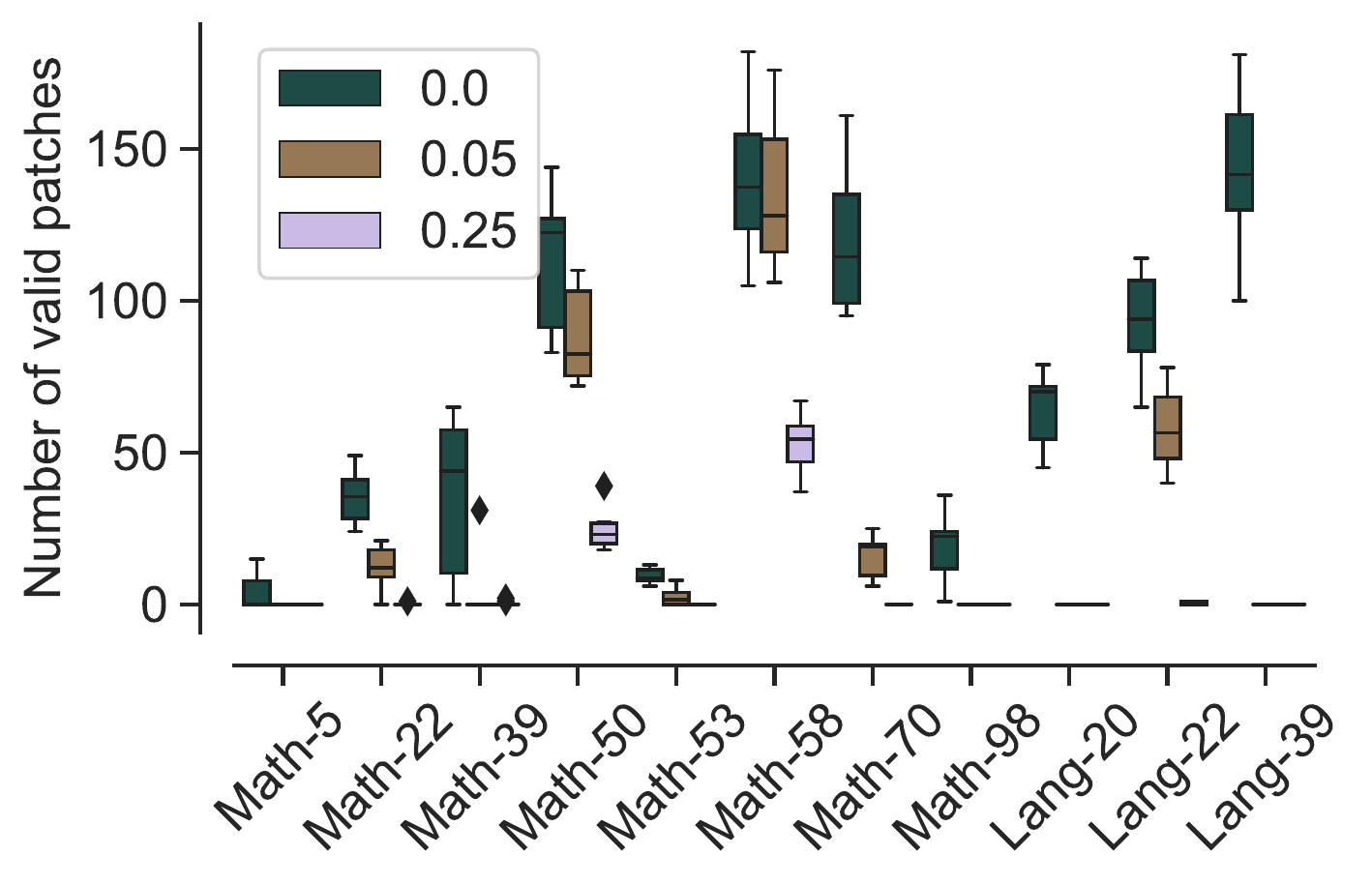}
    \caption{Impact of flakiness on the number of valid patches (showed as boxplots) generated by ARJA, for $p = 0$, $0.05$ and $0.25$. The effect of flakiness appears case-dependent and has disastrous consequences. This affects mainly the bugs that ARJA had a harder time fixing in the non-flaky variant. Repair is less effective as the flaky test failure rate increases.}
    \label{fig:results-arja}
\end{figure}

 We first run ARJA on each unmodified buggy program 10 times and analyze the obtained number of valid patches. Then, we repeat the same experiment by modifying the test suite of the buggy programs to introduce flaky tests, creating flaky variants of the programs. Since ARJA runs all tests before applying fault localization, we anticipate that making all tests of the buggy program flaky would have a very strong impact -- a fact later confirmed by the results of RQ4. Thus, to allow for relevant measurement, we render flaky only the tests belonging to the same test class as the initial failing tests, knowing that not all of them cover the likely-buggy statements. We assign each such test with a flakiness failure probability of 0.05 and perform 10 runs of ARJA on each flaky program. In another series, we change this probability to 0.25. This scenario simulates the application of ARJA to portions of the code with a high rate of flakiness.

We compare the number of the generated valid patches with and without flakiness. We do not consider genuine patches (i.e. semantically equivalent to the corresponding developers' patch) here because the randomness of the process would require intensive manual inspections of the patches produced across all runs.

Figure~\ref{fig:results-arja} shows, as boxplots and for each bug, the number of valid patches generated by ARJA for both the original (unmodified buggy) variant and the flaky variants. The impact of flakiness appears case-dependent. For instance, when we consider a $0.05$ probability of flakiness failures, the number of patches generated for Math-58 is barely reduced (median is $7\%$ lower). The reduction is more noticeable on Math-50 (-33\%) and Lang-22 (-40\%), although the total number remains high overall (>50 for all repetitions). In Math-70, the median number of valid patches decreases from 113 to 16 (-83\%). Still, ARJA produced valid patches at every run. Valid patches were also generated for the flaky variants of Math-22 (-66\%) and Math-53 (-82\%) in nine and seven runs out of ten, respectively. 

Flakiness has disastrous effects on the remaining projects. Math-98, Lang-20 and Lang-39 illustrate the case where ARJA generates patches at each run for the original (non-flaky) variant but could generate none for the flaky variant (-100\%). Math-5 and Math-39 are projects where ARJA may fail to generate patches even in the original variant, being successful only four and seven times, respectively. With flakiness, this number is reduced to zero in both cases (-100\%). Overall, while the impact of flakiness varies a lot from one buggy program to the other, the most negative scenarios tend to occur in programs for which ARJA could hardly generate a valid patch already in the non-flaky variants.

The number of (flaky) tests executed is also an important factor for incorrectly labelling patches (labelling as invalid). The programs of the Math project, for which the effectiveness of ARJA is the least affected, appear to be those that execute fewer tests (13 tests for Math-58, 7 tests for Math-70, 43 for Math-50). However, we found no strong correlation between the number of tests and the decrease in the number of valid patches. As an illustration, the median number of generated patches for Math-22 (81 executed tests, including 18 flaky tests) decreases from 36 to 12, while it decreases from 115 to 19 for Math-70 (7 executed tests, including 4 flaky tests). This means that other factors contribute significantly to this reduction, which we investigate in RQ3.

When increasing the probability of flaky test failures to $0.25$, ARJA can generate valid patches at each run for only two buggy programs (Math-50 and Math-58). It managed to produce one valid patch for Math-22 across the ten runs, three for Math-39, and none for the other projects. These results confirm the trend previously observed and indicate that applying repair methods is pointless above a certain degree of flakiness.\\

\begin{tcolorbox}[colback=white]
The decrease in effectiveness of non-deterministic repair due to flaky tests is case-specific, ranging from -7\% to -100\%. The effectiveness decreases more as the flaky test failures occur more frequently. 
\end{tcolorbox}

\subsection{RQ3: Targeted Fault Localization in Program Repair}

\begin{figure*}[t!]
\vspace{-0.5em}
\centering
\subfloat[Failing tests]{\includegraphics[width=0.33\textwidth]{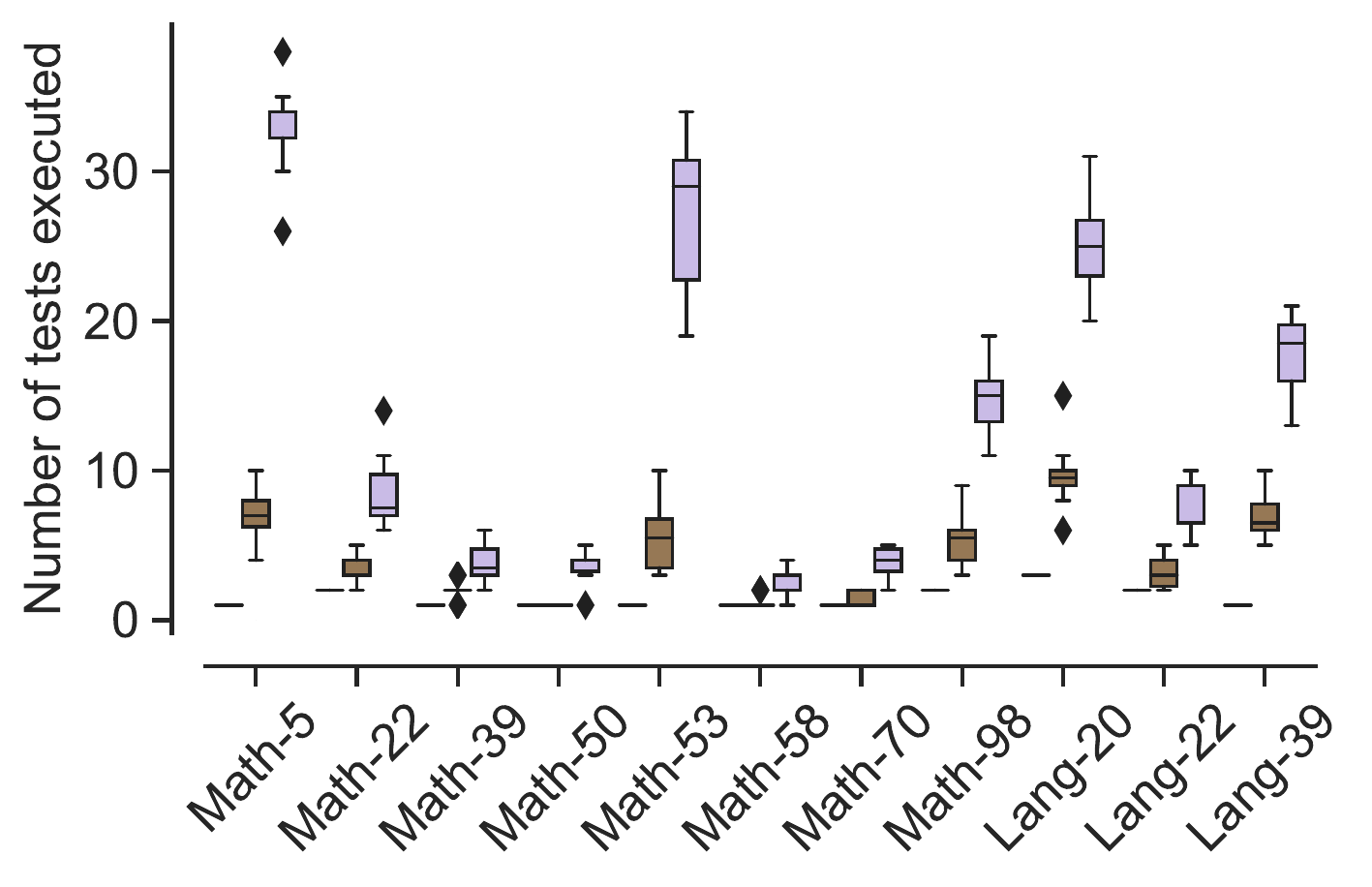}\label{fig:boxplot_depth}}
\subfloat[Positive tests]{\includegraphics[width=0.33\textwidth]{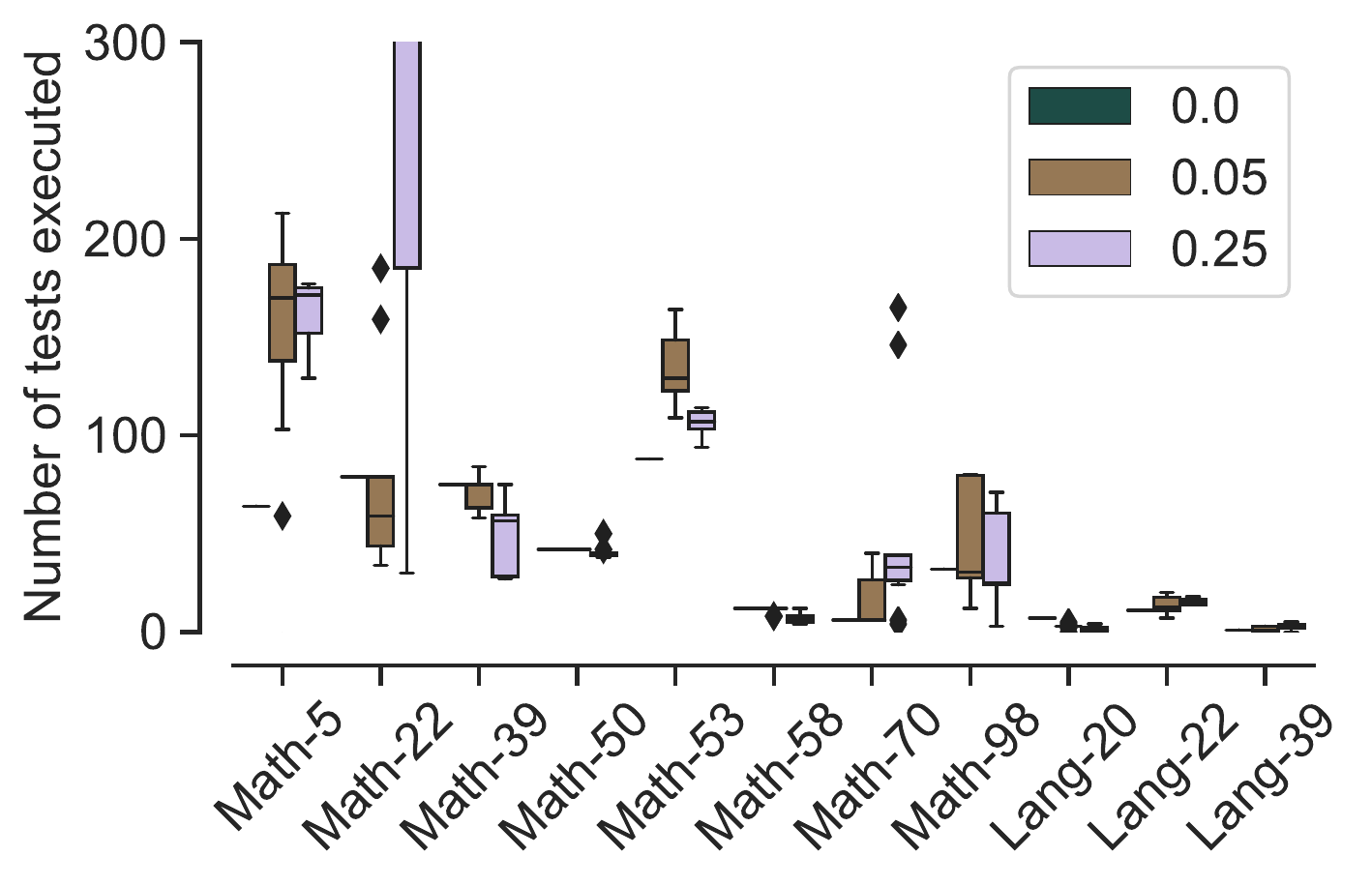}\label{fig:boxplot_connectivity}}
\subfloat[All tests]{\includegraphics[width=0.33\textwidth]{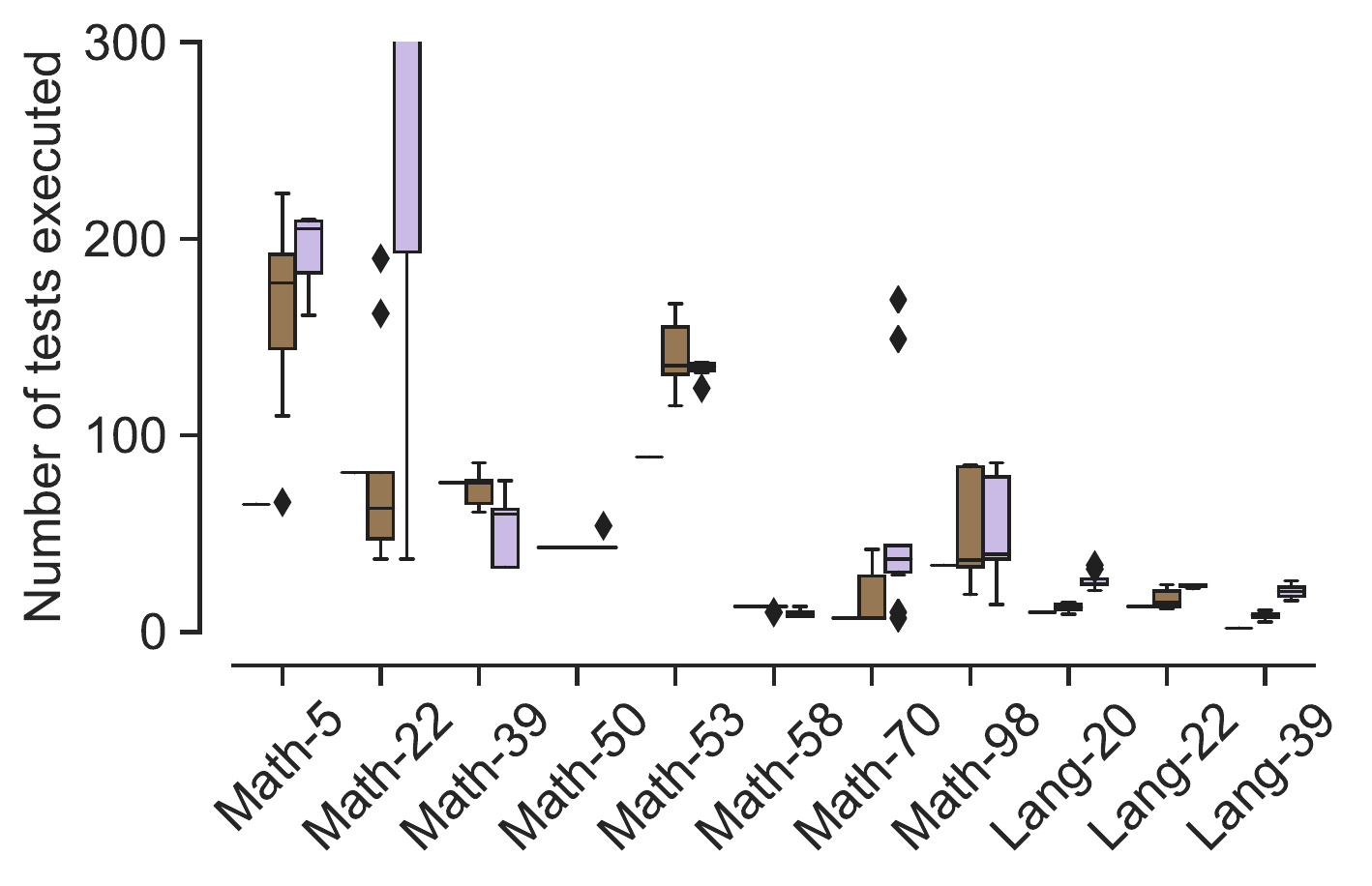}\label{fig:boxplot_depth_connectivity}}
\caption{The number of tests, failing, positive (passing tests that cover one or more suspicious statements) and all, executed by ARJA, for each program and over 10 runs, when using FlakiMe with $p = 0$, $0.05$ and $0.25$. Test flakiness creates discrepancies in the test results that are executed against candidate patches. This has a double effect; the waste of computational resources and a higher risk of ignoring important test signals.}  
\label{fig:results-test-number}
\end{figure*}

The failing and the positive tests (passing tests that cover one or more suspicious statements) identified during the fault localization step determine the set of ingredients to produce patches and are executed against any candidate patch. Their number is, thus, an indication of both the size of the search space and the risk of discarding a valid patch due to flakiness. Hence, we report on the number of failing and positive tests that were identified during our previous experiments on ARJA, in the flaky and non-flaky cases. 

Figure \ref{fig:results-test-number} shows the number of tests (failing and positive) executed by ARJA over the 10 different runs, for each buggy program. We observed that flakiness introduces random variations in these numbers. A higher probability of flaky test failure yields a higher number of failing tests (up to +900\% for $p = 0.05$, +3,700\% for $p = 0.25$). For $p = 0.05$, the total number of executed tests ranges from -54\% to +500\% of the real number (at $p = 0$); for $p = 0.25$, it ranges from -59\% to +2,314\%. This is because flaky tests that fail when applying fault localization introduces new suspicious statements which, in turn, makes it necessary to execute additional covering tests.  Surprisingly, we observe a mixed behaviour: more flakiness leads to a higher number of test executions (as expected) but it can also lower this number. A possible explanation for this is that Ochiai -- the suspiciousness formula used by default by ARJA -- depends on the total number of failing tests. Thus, the failing flaky tests may decrease the suspiciousness of the statements covered by the real failing tests. If their suspiciousness goes below the predefined threshold, ARJA ignores those statements and their covering tests. We investigate this phenomenon in more depth in RQ4.

We pursue our investigation by studying the practical benefits of making fault localization target the real failing test cases. We conduct controlled experiments where we compare the number of valid patches produced by ARJA in (1) the previous flaky case (with $p = 0.05$) where fault localization is applied as is (thus, considering all flaky tests that failed) and (2) a new case simulating the application of fault localization to the real failing test case only. To build this second case, we specify that only a real failing test is flaky, with a flakiness failure probability equivalent to the combined probability that any of the tests against which the patches are evaluated fails due to flakiness. Doing so allows discarding any suspicious statements, tests and ingredients that are artificially added (due to flakiness) at the fault localization step while keeping the same actual probability for a valid patch to fail. As before, we run ARJA 10 times on each variant of each buggy program and expect to observe improvements in the number of valid patches.

\begin{figure}
\vspace{-0.5em}
    \centering
    \includegraphics[width=0.75\linewidth]{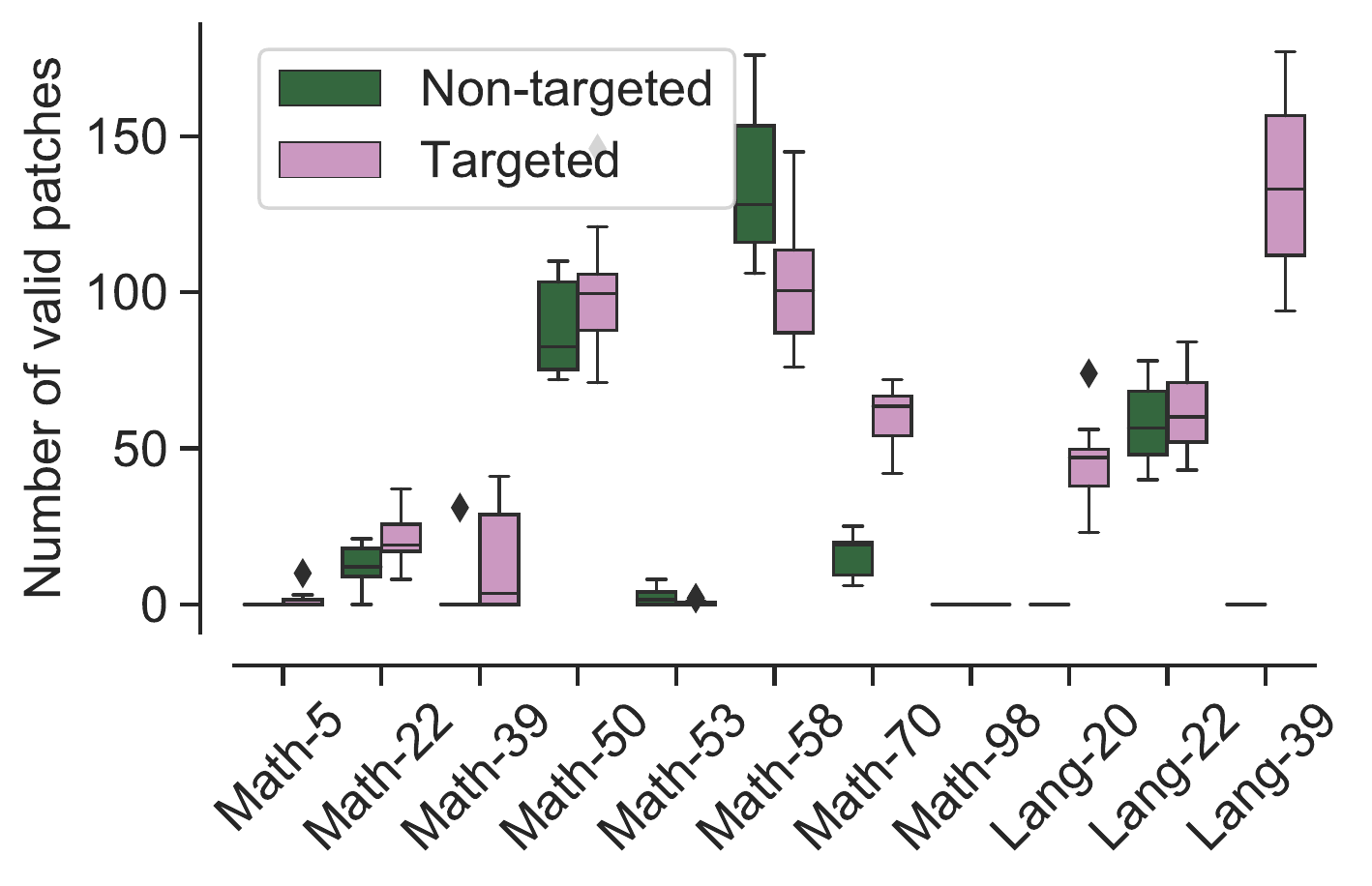}
    \caption{Number of generated valid patches obtained when applying fault localization on all tests (non-targeted) and only the real failing tests (targeted).}
    \label{fig:results-targeted}
    \vspace{-0.5em}
\end{figure}

Figure~\ref{fig:results-targeted} shows the number of generated valid patches in this new case (\texttt{Targeted}) and in the previous case (\texttt{Non-Targeted}), with a flakiness failure probability of 0.05. We observe a clear improvement in the targeted case, ARJA being even able to generate many valid patches for programs it could not repair in the non-targeted case. For the remaining programs, the median number of valid patches is up to +334\% higher. A Wilcoxon signed-rank test reveals that the differences are statistically significant, with a p-value of $1.86524 \times 10^{-06}$. 

The only exception to this are Math-50 and Math-58, for which we observe a decline. For Math-50, in both cases ARJA generate no valid patch in half of the runs, although the difference is statistically significant (p-value of $0.44$). For Math-58, the diminution of -21\% can be explained through a detailed look at the results: in the non-targeted case, adding a flaky failing test to the set of executed tests always results in reducing the number of positive tests. This means that the candidate patches are executed against fewer tests and can potentially be labelled as valid although they would make the missing tests fail. Overall, these results show the importance of identifying the failing test cases on which one should apply fault localization, to avoid corrupting both the patch search space and the validation process. \\

\begin{tcolorbox}[colback=white]
Applying fault localization on failing flaky tests corrupts the patch search space, removing valid patches and adding invalid ones. 
Targeting the real failing tests allows generating patches for four more programs and yields up to +334\% more valid patches. 
\end{tcolorbox}

\subsection{RQ4: Suspicious Statement Selection}

\begin{figure*}[t!]
\vspace{-0.5em}
\centering
\subfloat[Accuracy]{\includegraphics[width=0.3\textwidth]{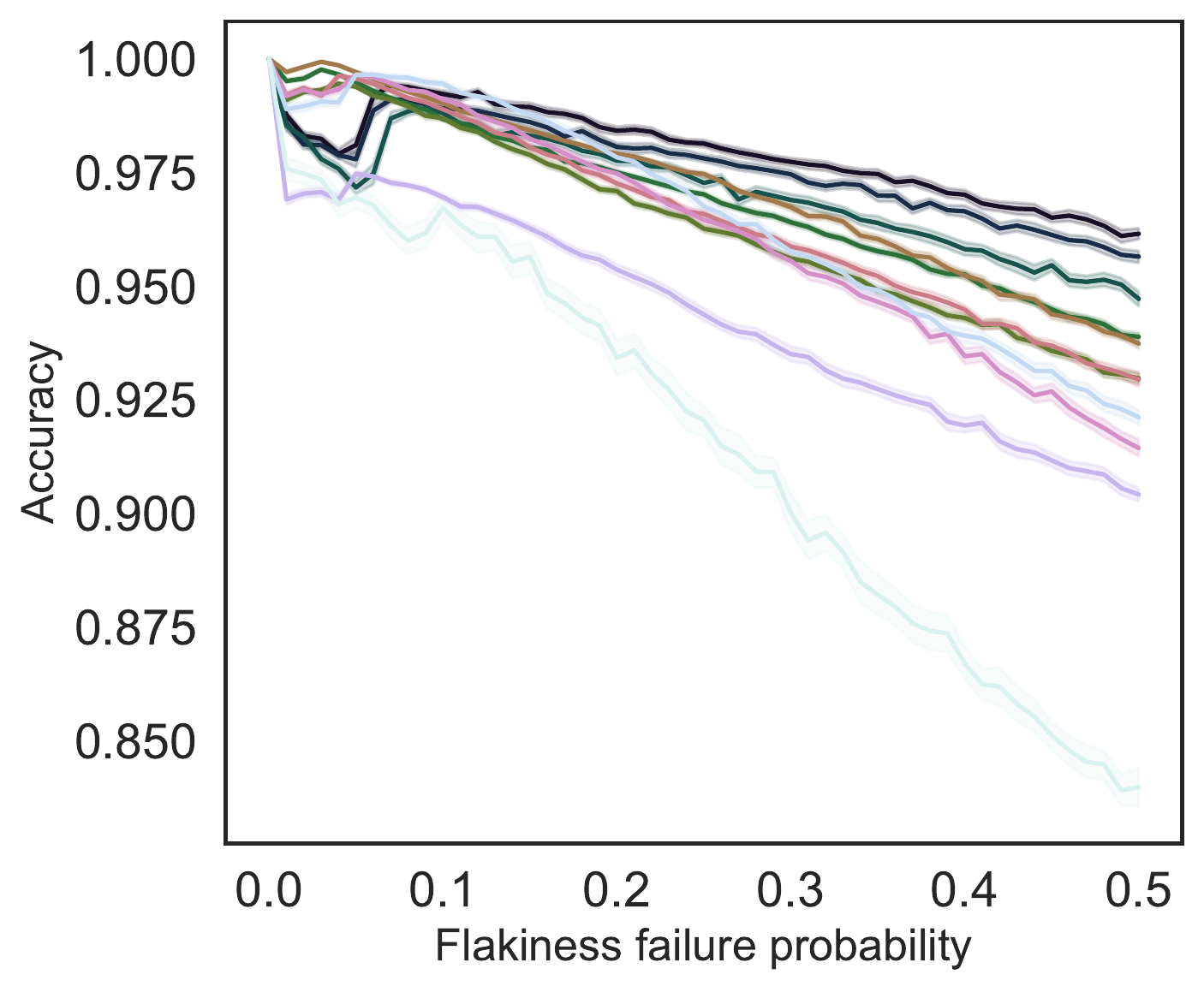}\label{fig:sbfl_accuracy}}
\subfloat[Precision]{\includegraphics[width=0.3\textwidth]{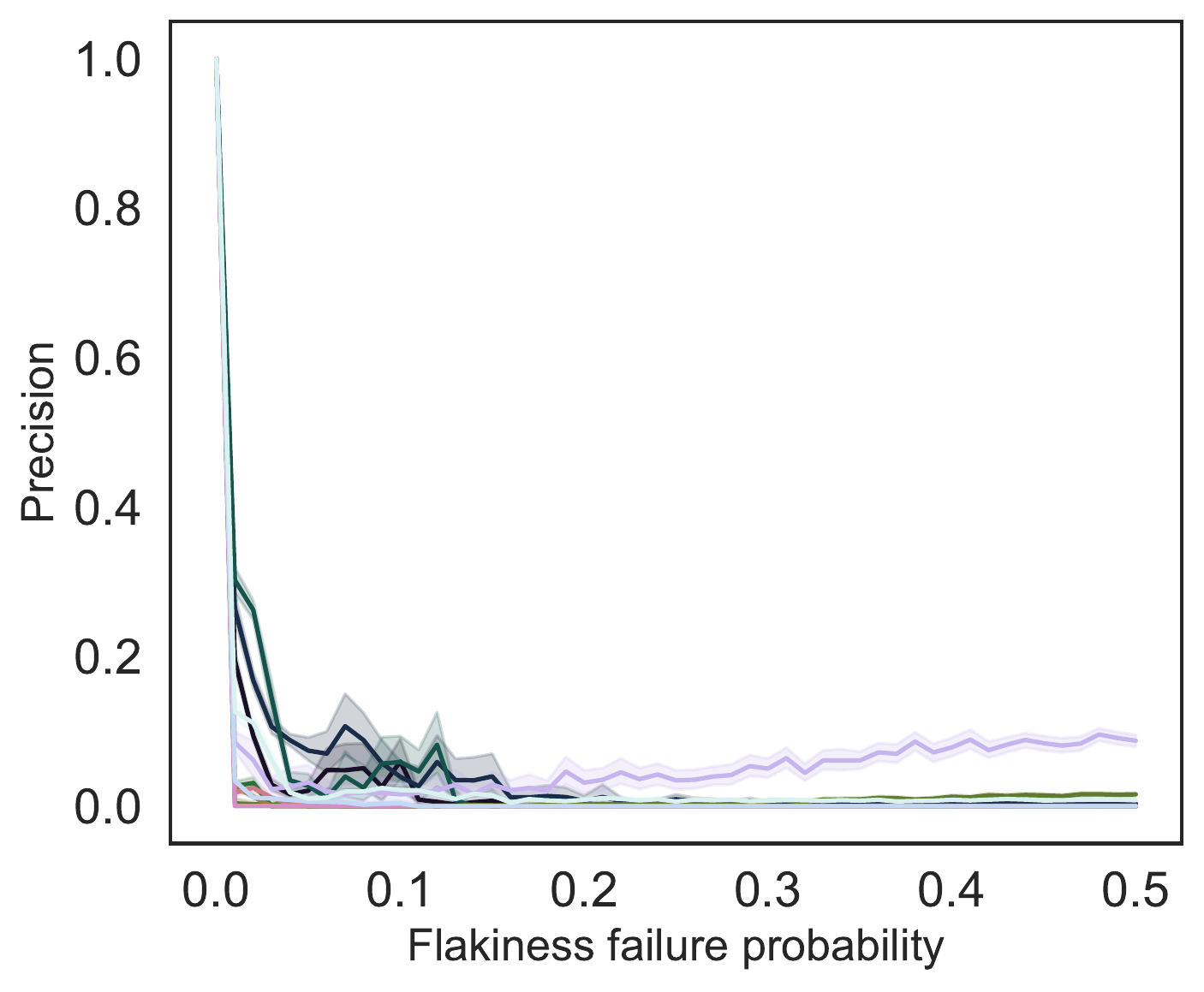}\label{fig:sbfl_precision}}
\subfloat[Recall]{\includegraphics[width=0.3\textwidth]{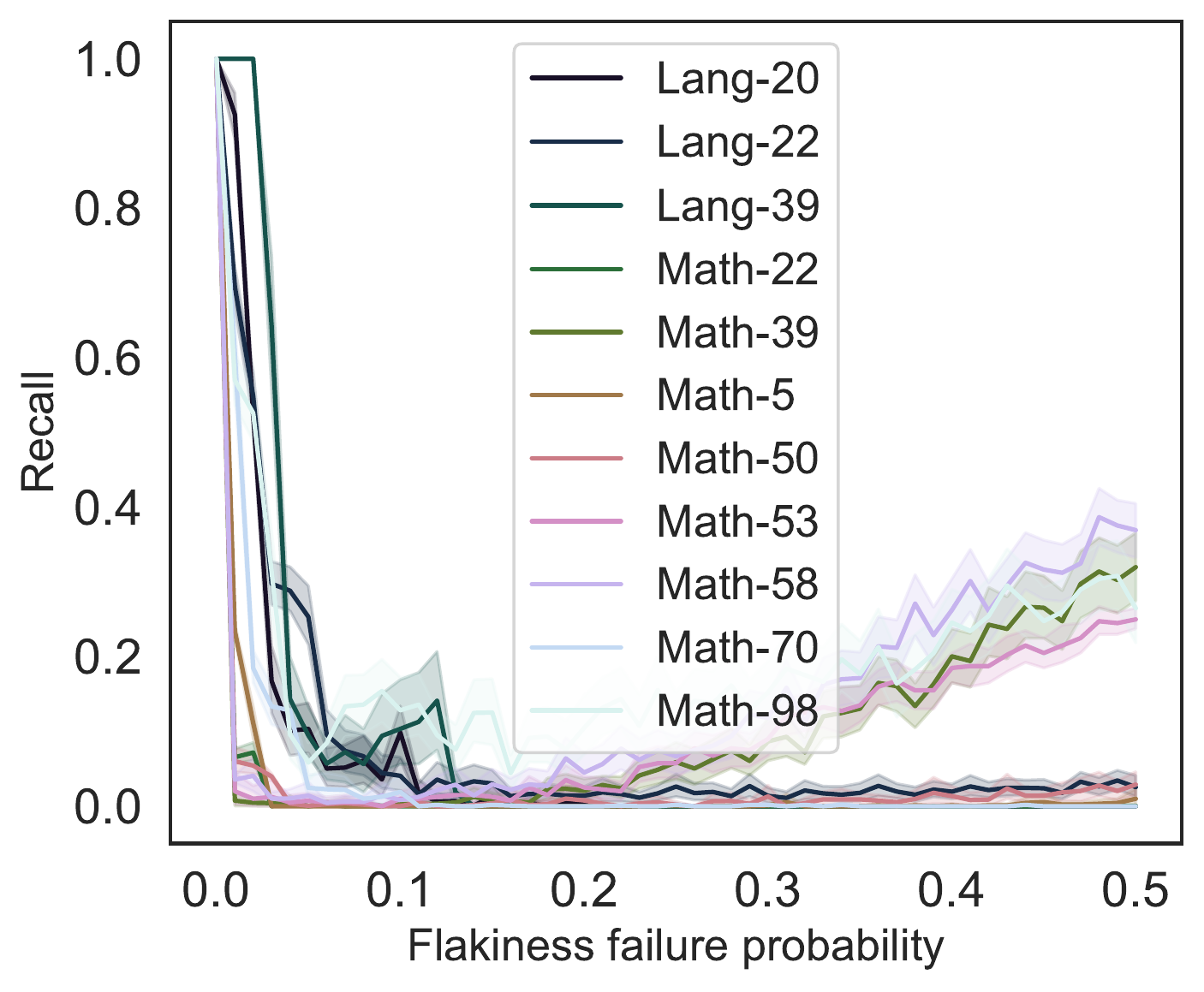}\label{fig:sbfl_recall}}
\caption{Suspicious statement selection when the flakiness failure probability increases.}  
\label{fig:sbfl_math}
\end{figure*}

To evaluate how the alterations in suspiciousness scores can impact the set of suspicious statements used by repair methods, we record, for each unmodified buggy program, the statements retained by ARJA after filtering. As recommended in the original paper \cite{Yuan2018}, we specify that all statements whose Ochiai score is below 0.1 are discarded. Taking the set of selected statements and discarded statements as the ground truth, we compute their counterparts in flaky variants of the program. Then, we define the robustness of Ochiai metrics against flakiness as its capability to preserve the original set of suspicious statements. 

We measure this robustness using the standard metrics of accuracy, precision and recall. Accuracy indicates the percentage of statements that remain in their class (selected or discarded), a coarse-grained view of how much the sets are altered. Precision measures the percentage of selected statements (in the flaky case) that indeed had to be selected (were selected in the non-flaky case). Thus, the lower the precision, the more the patch search space is increased with patches that do not target the real buggy statements. Recall measures the percentage of real suspicious statements that remained selected in the flaky case. A lower recall means a higher risk of discarding the real buggy statements.

We compute the accuracy, precision and recall for different flakiness failure probabilities, ranging from 0 (non-flaky case) to 0.5 and increasing by steps of $0.01$. For each probability value, we repeat the experiment 100 times. Note that our goal here is to assess the overall effect of flakiness on fault localization. Thus, unlike the previous experiments where we flaked only the test classes containing the real failing tests, we make flaky tests of the buggy program.  

Figure~\ref{fig:sbfl_math} shows the trend followed by the accuracy, precision and recall of the suspicious statement selection when the probability of flakiness failure increases. In Figure~\ref{fig:sbfl_accuracy}, we observe that the accuracy tends to decrease linearly with a higher rate of flakiness failures but, overall, remains high for all projects (over 90\% for a flakiness failure probability of 0.50). Indeed, most statements are not covered by any failing test in the non-flaky program and remain so in the flaky cases. As the degree of flakiness increases, the number of errors (false positives and false negatives) naturally increases.


Surprisingly, both precision and recall values drop and become close to zero as soon as small degrees of flakiness ($p < 0.05$) are introduced. This decline is explained by the fact that the number of true positives (selected statements that are indeed the right ones) form a tiny percentage compared to the number of true negatives (non-selected statements that are not the right ones). Flaky tests increase the number of failing tests $n_f$, which in turn introduces additional suspicious statements. Because of that, the suspiciousness score of the selected statements drops below the 0.1 threshold, which justifies the weak recall (more false negatives). 

Similarly we see that few cases become false positives (selected statements that are not initially suspicious). Now it is interesting to see that if we combine these cases with the small rate of true positives we get a precision decline. As flakiness becomes more apparent, the precision does not change but the recall improves slightly. This is because the failing tests have an increasing number of overlapping statements, which leads to an overall increase of their suspiciousness score and, thus, to more (true and false) positive cases and fewer false negative cases. 

Overall, the slightest degrees of flakiness (i.e., $p = 0.01$) can disrupt the threshold-based suspicious statement selection by a 90\% drop in precision and recall.  This shows that the adopted threshold of the  threshold-based fault localization is yet another factor contributing to ARJA's loss of effectiveness. As shown by our results, the potential benefits of this threshold (reducing the number of tests to execute) must be balanced with the risk of executing flakiness, which can dramatically reduce the performance of program repair. Lowering the threshold may help, but still necessitates a clear a priory knowledge of the particular flakiness failure rate.\\

\begin{tcolorbox}[colback=white]
Even $1\%$ of flakiness is sufficient to disrupt threshold-based suspicious statement selection. We found that both precision and recall values drop by an amount higher than 90\%. Without user feedback the fault localization cannot target real failing tests, so the use of the threshold should be avoided. 
\end{tcolorbox}




\subsection{Threats to Validity}
\label{sec:threats}

The major threat to the validity of our results is the way we simulated the impact of flakiness on test results. First, we assume that the techniques are applied in a black-box way, without knowing the location of the flaky tests. While this is a realistic setting, incorporating project-specific knowledge could help observing special cases and perhaps less pessimistic results. Moreover, the likelihood of flakiness is known to increase when particular anti-patterns occur \cite{Palomba2019}, an ignored fact that may induce  case specific distributions. While this may change the magnitude of the findings per case, it cannot change the key considerations drawn out by the use of FlakiMe. 
Nevertheless, FlakiMe can support a plethora of scenarios, such as the above ones, providing experimental control on flakiness. We therefore, expect that future work will further alleviate such threats by considering test suites' and projects characteristics. 

Given that FlakiMe alters the execution of tests (and not the source code), we cannot systematically control the execution flow of the program under test. Moreover, we make flaky tests fail at the end of their execution. Therefore, no test that flakes, changes its coverage. Nevertheless, we do not consider this as a major threat since previous studies have shown that such phenomenon is infrequent ~\cite{Shi2019}. 


Regarding PRAPR, our analysis targets the long-term behaviour of the tool when facing identically distributed flakiness failures. Running FlakiMe on real-world flaky tests could lead to different results because of random variations and non-uniform distribution of the flaky tests. Still, our conclusions are aligned with previous studies stating that the effectiveness of deterministic approaches decreases as the number of failing tests increases \cite{Kong2018}. 




%% file: 7-conclusion.tex
\section{Conclusion}
\label{sec:conclusion}

We presented a test flakiness tool, FlakiMe, that allows researchers experiment with laboratory-controlled test flakiness. 
FlakiMe is customizable and can (easily) simulate a wide range of conditions and scenarios. We used FlakiMe to perform two case studies and demonstrate the opportunities and benefits that it brings. 
Interestingly, we showed that putting flakiness under laboratory control adds a new dimension to software testing studies, which is the simulation of  a world where ``All Tests are Flaky'' \cite{mhpoh:scam18-keynote} or at least we should consider all as {\em potentially} flaky. Such a world allows establishing a better understanding of the effects of flakiness and paves the way for developing robust (on flaky tests) test techniques. 

We demonstrated that mutation, a popular test assessment metric, is impacted by flaky tests, i.e., mutation score is inflated by approximately 5\%-10\% depending on the level of flakiness. This effect is however small as the introduced noise is similar among all cases making the metric relatively stable, it varies from 2\% to 4\%. 

In program repair, our results showed that the fault localization step is particularly sensitive to test flakiness. Such sensitivity can have disastrous effects on patch generation. Thus, to make program repair techniques robust against flaky tests, one should revisit the key decisions and assumptions made during fault localization.

For example, in a scenario where some `real' failing tests are specified (by the user), a tailored fault localization procedure that considers only these tests helps to prevent an artificial increase of the patch search space as well as useless runs of the candidate patches with flaky test cases. 

Based on what we have seen, flakiness may also modify the suspiciousness score of the target statements (candidate statements for mutation). If this information is used only for prioritization (without discarding statements), the program repair would still generate the same number of valid patches; only their order would change. On the contrary, if it is used for selection/filtering (discarding less suspicious statements), the program repair would consider more non-buggy statements -- resulting in losses of efficiency -- and, in extreme cases, the buggy statements may be skipped -- resulting in the impossibility to generate valid patches. Thus, one should balance the gain in efficiency achieved by filtering procedures against those risks induced by test flakiness. 

Nevertheless, a valid patch has more risk to be discarded when it is covered by more (potentially flaky) tests. To alleviate this effect, approaches based on genetic programming can include the number of covering tests into their fitness function. Doing this allows them to direct the search towards the patches covered by fewer tests. Of course, achieving good results would require well-tuned parameterization to set appropriately the weight of each constituent of the fitness function. Still, our experiments on PRAPR have shown that genuine patches are, on average, covered by fewer tests than the other valid patches, which supports the above point.